\documentclass{aa}  
\usepackage[english]{babel}
\usepackage{graphicx}
\usepackage{comment}
\usepackage[a4paper]{geometry}
\usepackage{xcolor}

\usepackage{amsmath}
\usepackage[colorlinks=true, allcolors=blue]{hyperref}
\newcommand{\healpix}{\texttt{HEALPix}}
\newcommand{\nnhealpix}{\texttt{NNhealpix}}
\newcommand{\srolltwo}{\texttt{SRoll2}}
\newcommand{\planck}{\textit{Planck}}
\newcommand{\nside}{{N_{\rm side}}}

\title{Inference of the optical depth to reionization $\tau$ from \planck{} CMB maps with convolutional neural networks}
\titlerunning{Inference of $\tau$ from \planck{} maps with convolutional NNs}

\author{\small 
Kevin Wolz\inst{1,2}~\thanks{email: K.~Wolz, kevin.wolz93@gmail.com}
\and
Nicoletta Krachmalnicoff\inst{1,2,3}~\thanks{email: N.~Krachmalnicoff, nkrach@sissa.it}
\and
Luca Pagano\inst{4,5}~\thanks{email: L.~Pagano, luca.pagano@unife.it}
}

\institute{\small
International School for Advanced Studies (SISSA), Via Bonomea 265, I-34136 Trieste, Italy \goodbreak
\and
National Institute for Nuclear Physics (INFN) -- Sezione di Trieste, Via Valerio 2, I-34127 Trieste, Italy \goodbreak
\and
Institute for Fundamental Physics of the Universe (IFPU), Via Beirut 2, I-34151 Grignano (TS), Italy \goodbreak
\and
Dipartimento di Fisica e Scienze della Terra, Universit\`a degli Studi di Ferrara and INFN -- Sezione di Ferrara, Via Saragat 1, I-44122 Ferrara, Italy \goodbreak
\and
Université Paris-Saclay, CNRS, Institut d’Astrophysique Spatiale, F-91405 Orsay, France\goodbreak
}

\begin{document}

\abstract
{
{The optical depth to reionization, $\tau$, is the least constrained parameter of the cosmological $\Lambda$ cold dark matter ($\Lambda$CDM) model. To date, its most precise value is inferred from large-scale polarized cosmic microwave background (CMB) power spectra from the High Frequency Instrument (HFI) aboard the \planck{} satellite. These maps are known to contain significant contamination by residual non-Gaussian systematic effects, which are hard to model analytically. Therefore, robust constraints on $\tau$ are currently obtained through an empirical cross-spectrum likelihood built from simulations. \\
In this paper, we present a likelihood-free inference of $\tau$ from polarized \planck{} HFI maps which, for the first time, is fully based on neural networks (NNs). NNs have the advantage of not requiring an analytical description of the data and can be trained on state-of-the-art simulations, combining the information from multiple channels.\\
By using  Gaussian sky simulations and \planck{} \srolltwo{} simulations, including CMB, noise, and residual instrumental systematic effects, we trained, tested, and validated NN models considering different setups. We inferred the value of $\tau$ directly from Stokes $Q$ and $U$ maps at $\sim 4^\circ$ pixel resolution, without computing angular power spectra.\\
On \planck{} data, we obtained $\tau_{\rm NN}=0.0579\pm 0.0082$, which is compatible with current $EE$ cross-spectrum results but with a $\sim30\%$ larger uncertainty, which can be assigned to the inherent nonoptimality of our estimator and to the retraining procedure applied to avoid biases. \\
While this paper does not improve on current cosmological constraints on $\tau$, our analysis represents a first robust application of NN-based inference on real data, and highlights its potential as a promising tool for complementary analysis of near-future CMB experiments, also in view of the ongoing challenge to achieve the first detection of primordial gravitational waves.
}}

\maketitle

\section{Introduction}
Cosmic reionization, the period in cosmic history that accompanies the ignition of the first stars, is of great interest to both astrophysics and cosmology. At recombination, about 380,000 years after the big bang, free electrons were bound in hydrogen atoms, causing the decoupling of matter from the photon field that we observe today as the cosmic microwave background (CMB). This is when the Universe entered the electrically neutral phase, called ``cosmic dark ages.'' It is presumed that about 200 million years later, cold hydrogen gas had collapsed gravitationally in dark matter halos, forming the first stars. These earliest compact sources of UV radiation heated up the surrounding hydrogen gas, progressively ionizing the whole Universe via bubbles of expanding HII regions. The so-called Gunn-Petersen trough \citep{1965ApJ...142.1633G, 1965Natur.207..963S} in the absorption spectrum of high-redshift quasars indicates the presence of neutral hydrogen in the intergalactic medium (IGM) along the corresponding lines of sight. The detection of this feature in the spectra of some $z>5.8$ quasars by the \textit{Sloan Digital Sky Survey} provided the first spectroscopic evidence for reionization \citep{SDSS:2000jpb, SDSS:2001tew, SDSS:2001emm}. Modern quasar measurements indicate that the epoch of reionization was completed by $z\approx 5.3$ \citep{Qin:2021gkn, Villasenor:2021ksg, Bosman:2021oom}.

Reionization plays a crucial role in cosmology as well. Photons that are emitted during recombination have a finite probability of Compton scattering with free electrons released during reionization. For us as observers, this has two effects: firstly, CMB photons traversing the IGM cause a uniform damping of CMB anisotropies at scales below the cosmological horizon at the epoch of reionization ($\ell>20$). Secondly, a statistically relevant fraction of CMB photons scatter into our line of sight, carrying a nonzero net polarization observable as secondary anisotropies in the CMB polarization. The first effect reduces the CMB power spectrum amplitude of both unpolarized and polarized components by a factor $e^{-2\tau}$, where $\tau$ is the optical depth to reionization, defined as
\begin{equation}\label{eq:tau}
    \tau = \int_{t(z_{\rm CMB})}^{t_0} n_{\rm e} \sigma_{\rm T}\, c\,{\rm d}t' \, .
\end{equation}
Here, $z_{\rm CMB}\approx 1100$ is the time of last scattering between photons and baryons, $n_{\rm e}$ is the electron number density, $\sigma_{\rm T}$ is the Thomson scattering cross section, and $c$ is the speed of light. The second effect introduces large-scale power in the polarized CMB proportional to $\tau^2$, affecting scales larger than the horizon at the epoch of reionization ($\ell<20$). Full-sky space missions such as {\it WMAP} and \planck{} have been able to measure this ``reionization bump'' through pixel-based and power-spectrum-based analysis methods. The {\it WMAP} nine-year data release cites $\tau=0.089\pm0.014$ \citep{2013ApJS..208...19H}, a value that later turned out to be biased high due to Galactic dust emission \citep{Planck:2015bpv,Natale:2020owc}. \planck's low-frequency instrument (LFI) polarization data at 70 GHz contain less large-scale systematics than the High Frequency Instrument (HFI) data at 100~GHz and 143~GHz, motivating the \planck{} Collaboration to perform map-based analysis on LFI data and cross-spectrum analysis on HFI data. The \planck{} 2018 legacy release cites $\tau=0.063\pm0.020$ as inferred from LFI data and $0.051\pm0.009$ from HFI data \citep{Planck:2019nip}. The cross-spectrum analysis method of \planck{} HFI data at 143~GHz and 100~GHz yields the tightest constraint to date, while avoiding the bias arising from uncorrelated noise in the individual frequency channels.

The \planck{} 2018 legacy polarization data products at large scales are known to be affected by residual contamination from  instrumental systematic effects. As investigated in \cite{2014A&A...571A...6P} and \cite{Delouis:2019bub}, the most important effects at 143~GHz and 100~GHz are temperature-to-polarization ($T$-to-$P$) leakage due the analog-to-digital converter nonlinearity (ADCNL), uncertainties on the detectors' orientation and polarization efficiencies, $T$-to-$P$ leakage due to bandpass mismatch and inaccurate Galactic foreground modeling, and a varying time constant associated with the heat transfer to the bolometers. In general, these systematic effects follow non-Gaussian statistical distributions and are expected to correlate among different channels, mainly because they are partially sourced by the temperature signal. Several updated mapmaking codes have been published that improve on the systematics cleaning, such as \srolltwo{} \citep{Delouis:2019bub}, and {\tt NPIPE} \citep{Planck:2020olo}. The \srolltwo{} algorithm, an upgraded version of the \planck{} Collaboration's {\tt SRoll} algorithm \citep{2016A&A...596A.107P}, iteratively cleans systematics from \planck's time-ordered data products. Major improvements in \srolltwo{} encompass a new gain calibration model that accounts for second-order ADCNL, updated foreground templates, and an internal marginalization over the polarization angles and efficiencies for each bolometer. 
The \srolltwo{} data products contain a significantly lower level of spurious systematic effects and a dipole residual power reduced by 50\% with respect to the \planck{} 2018 legacy data, falling below the noise level. The \srolltwo{} $EE$ cross-spectrum is dominated by the cosmological signal at all scales that were considered in the analysis ($2<\ell<30$).

In spite of the improved cleaning, a small residual contamination remains (mainly due to the second-order ADCNL effect), which may bias cosmological analyses. For their $100\times143$ GHz $EE$ cross-spectrum analysis of the \srolltwo{} data products, \cite{2020A&A...635A..99P} use an empirical likelihood built from realistic simulations \citep{Planck:2019nip,2020FrP.....8...15G}, motivating their choice by the expected non-Gaussianity of the maps and by the difficulty to model residual systematic effects analytically. They obtain $\tau=0.0566^{+0.0053}_{-0.0062}$ (68\% CL) from $EE$ only and $\tau=0.059\pm0.006$ when combining with $TT$ data. Compared with the $EE$ results from the \planck{} 2018 legacy release ($\tau=0.051\pm0.009$), this reduces the uncertainty by $\sim 40\%$ and increases the best-fit $\tau$ value by up to 0.9$\sigma$. More recently, \cite{deBelsunce:2021mec} applied various likelihood approximation schemes on $EE$ cross-spectrum data from \srolltwo{} maps, finding results compatible with \cite{2020A&A...635A..99P}, though slightly larger by $0.3\sigma$.\\

In recent years, neural network (NN)-based approaches to likelihood-free inference underwent a rapid development in cosmology, showing potential as an alternative tool for parameter estimation that does not require the existence of an analytical description of the data, but only relies on numerical simulations to train a regression model. In the general context of cosmology, a variety of machine learning (ML) techniques have been exploited and tested in recent years. Promising tools are being developed for many applications: from cosmic large-scale structure (LSS) simulations \citep{2022ApJS..259...61V}, to CMB lensing reconstruction \citep{Caldeira:2018ojb}, kinetic SZ detection \citep{Tanimura:2022fde}, or modeling and cleaning of Galactic foregrounds \citep{Jeffrey:2021fcg, Wang:2022ybb, Casas:2022teu, Krachmalnicoff_2021}. NN-based inference of cosmological parameters has seen significant progress in the context of observations of the LSS, where the complexity of the cosmological and astrophysical signals, together with the difficulty in the definition of optimal summary statistics, challenge analytical methods. Up to now, this approach has been tested on simulations \citep[see e.g.,][]{2022ApJS..259...61V}, with applications on real data still limited in number, although leading to promising results \citep[e.g.,][]{Fluri:2019qtp}. In this context, CMB data analysis could also benefit from the application of NN-based inference, helping overcome the limitations of traditional methods. This is relevant, for example, for the estimation of parameters affecting the large angular scales, such as the optical depth to reionization, which is critically hampered by the presence of spurious non-Gaussian signals, as outlined above.

This paper represents the first map-level cosmological inference on CMB data that is entirely based on convolutional neural networks (CNNs). We use CNNs to infer the optical depth to reionization $\tau$ and its statistical uncertainty from \planck{} multifrequency maps on the 100 and 143~GHz channels at scales $\gtrsim 4^\circ$, having trained and validated our findings on the \srolltwo{} simulations. Using moment networks \citep{Jeffrey:2020itg}, we infer $\tau$ and its statistical uncertainty $\sigma(\tau)$ from a single data set. In particular, we demonstrate:
\begin{enumerate}
    \item When training the CNN on simulations with realistic, correlated Gaussian noise, we achieve unbiased estimates of $\tau$ from maps.
    \item Our NN models can effectively combine multifrequency information, recognizing common features across channels, not only to reduce statistical uncertainty but also to diminish the impact of noise and systematic effects.
    \item Training on non-Gaussian data is necessary to obtain unbiased results on the \srolltwo{} test simulations and \planck{} data. Limited by a low number of simulations that contain \planck{} systematics, we were forced to build a retrained model, which increased the error bar on $\tau$ by $\sim 30\%$ in exchange for unbiased results.
\end{enumerate}

This paper is structured as follows. We present the simulations and data used in this work in Sect.~\ref{sec:simulations_data}, followed by the neural network inference method in Sect.~\ref{sec:nn_inference}. In order to validate this method, we apply it to a series of simulations and present the results in Sect.~\ref{sec:results_sims}. We discuss the final results on the \planck{} \srolltwo{} maps follows in Sect.~\ref{sec:planck} and conclude in Sect.~\ref{sec:conclusion}.

\section{Simulations and data}\label{sec:simulations_data}
The goal of our analysis is to build a NN model able to infer the value of the cosmological parameter $\tau$ from \planck{} low-resolution polarization input maps. In particular, in this work we used the \srolltwo{} maps at 100 and 143~GHz. To achieve this, we needed a large number of simulations to perform NN training, testing, and validation. We generated simulated maps that include CMB emission, noise, and instrumental systematic effects, as well as possible spurious signals coming from our Galaxy. In this section, we describe the simulations, the data, and the sky masks needed to avoid the highly contaminated Galactic plane region.

\subsection{Simulated CMB maps}

Polarized CMB anisotropies, observed at the \planck{} noise levels, can be sufficiently well represented by a spin-2 field with Gaussian statistics \citep{Planck:2019kim}. The $TT$, $TE$, and $EE$ power spectra characterize the probability distribution of CMB temperature and polarization anisotropies and can be described by the six parameters of the $\Lambda$ cold dark matter ($\Lambda$CDM) model. Analyses of small-scale temperature data from the \planck{} 2018 legacy release place a $0.5\%$ constraint on the parameter combination $10^9\,A_s\,e^{-2\tau}=(1.88\pm0.01)$ \citep{Planck:2018vyg}. Varying the two parameters $(A_s,\,\tau)$ simultaneously conditioned on $10^9\,A_s\,e^{-2\tau}=1.884$, coherent with previous studies \citep{2016A&A...596A.107P,Planck:2019nip,2020A&A...635A..99P,Planck:2020olo}, we used the Boltzmann solver {\sc CAMB}\footnote{\url{http://camb.info}} \citep{Lewis:1999bs} to generate a lookup table of $EE$ power spectra computed with the $\Lambda$CDM model. To build the simulated CMB maps used to train and validate our NN models, we discretized $\tau\in[0.01,\,0.13]$ with step size $\Delta\tau=5\times10^{-4}$. Since the other $\Lambda$CDM parameters have no substantial impact on polarized CMB spectra at low multipoles, we fixed them to the \planck{} 2018 legacy best-fit values $H_0=67.32$ km/s/Mpc, $\Omega_b h^2=0.02237$, $\Omega_c h^2=0.1201$, $n_s=0.9651$, $m_\nu=0.06$. From the tabulated power spectra, we uniformly drew 200,000 samples based on which we generated 200,000 pairs of full-sky Stokes $Q$ and $U$ maps using the \healpix{} package \citep{2005ApJ...622..759G}. We fixed the $Q$ and $U$ maps' angular pixel resolution by choosing $\nside=16$ (or a pixel size of $\sim4^\circ$)\footnote{A \healpix{} map has $N_{\rm pix} = 12N^{2}_{\rm side}$ pixels of the same area $\Omega_{\rm pix} = \pi/(3N^{2}_{\rm side}).$} and smooth each map with a cosine beam window function \citep{2009MNRAS.400..219B}, in analogy with the procedure used to generate the \planck{} \srolltwo{} maps (see Sect.~\ref{ssec:planck_maps}). These large scales retained in our maps correspond to multipoles $\ell\lesssim 50$, where the reionization bump leaves an observable imprint in the CMB $EE$ spectrum.

\subsection{Simulated Gaussian noise}\label{ssec:noise_sims}

\planck{} maps contain Gaussian instrumental noise which, in pixel space, is well described by the FFP8 covariance matrices \citep{2016A&A...594A..12P}. We drew samples from them for the \planck{} 100 and 143~GHz polarization channels \citep{2014A&A...571A...6P, 2016A&A...594A..13P}, obtaining 200,000 Gaussian noise maps at pixel resolution $\nside=16$ for both channels, respectively. We coadded the training maps of CMB and noise to obtain 200,000 \planck-like simulations on the full sky, out of which we selected 190,000 for training and 10,000 for validation. For the testing phase, we drew new noise samples in the same fashion as before, but coadded CMB simulations with fixed input values $\tau=0.05$, 0.06, and 0.07 and different seeds than the ones used for training and validation. In this way, we obtained three sets of 10,000 independent Gaussian test simulations with the fixed input cosmologies.

\subsection{\srolltwo{} simulations}\label{ssec:sroll2_sims}

The \srolltwo{} simulations \citep{Delouis:2019bub} improve on the {\tt SRoll} simulations published along with \planck’s third data release \citep{Planck:2020olo}. They are the result of applying the \srolltwo{} cleaning algorithm to a set of 500 \planck-like realistic sky simulations containing modeled noise, foregrounds, and instrument systematics. We chose the \srolltwo{} simulations as our reference for systematic effects present in the \srolltwo{} \planck{} data. All simulated maps are cleaned from Galactic foregrounds through a template fitting procedure, as described in \cite{2020A&A...635A..99P}. In order to produce our training set, we started with  400 out of the 500 original \srolltwo{} simulations containing pairs of $Q$ and $U$ full-sky maps at pixel resolution $\nside=16$ and two channels corresponding to 100~GHz and 143~GHz. 
To augment our original \srolltwo{} simulation set, we randomly drew \srolltwo{} 100~GHz and 143~GHz maps from the original 400 maps (with repetition), keeping corresponding $Q$ and $U$ maps together. This allowed us to assemble a total of 200,000 \srolltwo{} simulations. After coadding them with CMB simulations, we obtained a set of 200,000 polarized full-sky simulations, used for training and validation. For the testing phase, we made $3\times100$ copies of $100$ unseen \srolltwo{} maps and coadded them with $10,000$ CMB maps with fixed input $\tau=0.05$, $0.06$, and $0.07$, respectively. In this way we obtained a set of $3\times 10,000$ full-sky \srolltwo{} test simulations.

\subsection{\planck{} maps}\label{ssec:planck_maps}

\begin{figure}
    \centering
    \includegraphics[width=\columnwidth]{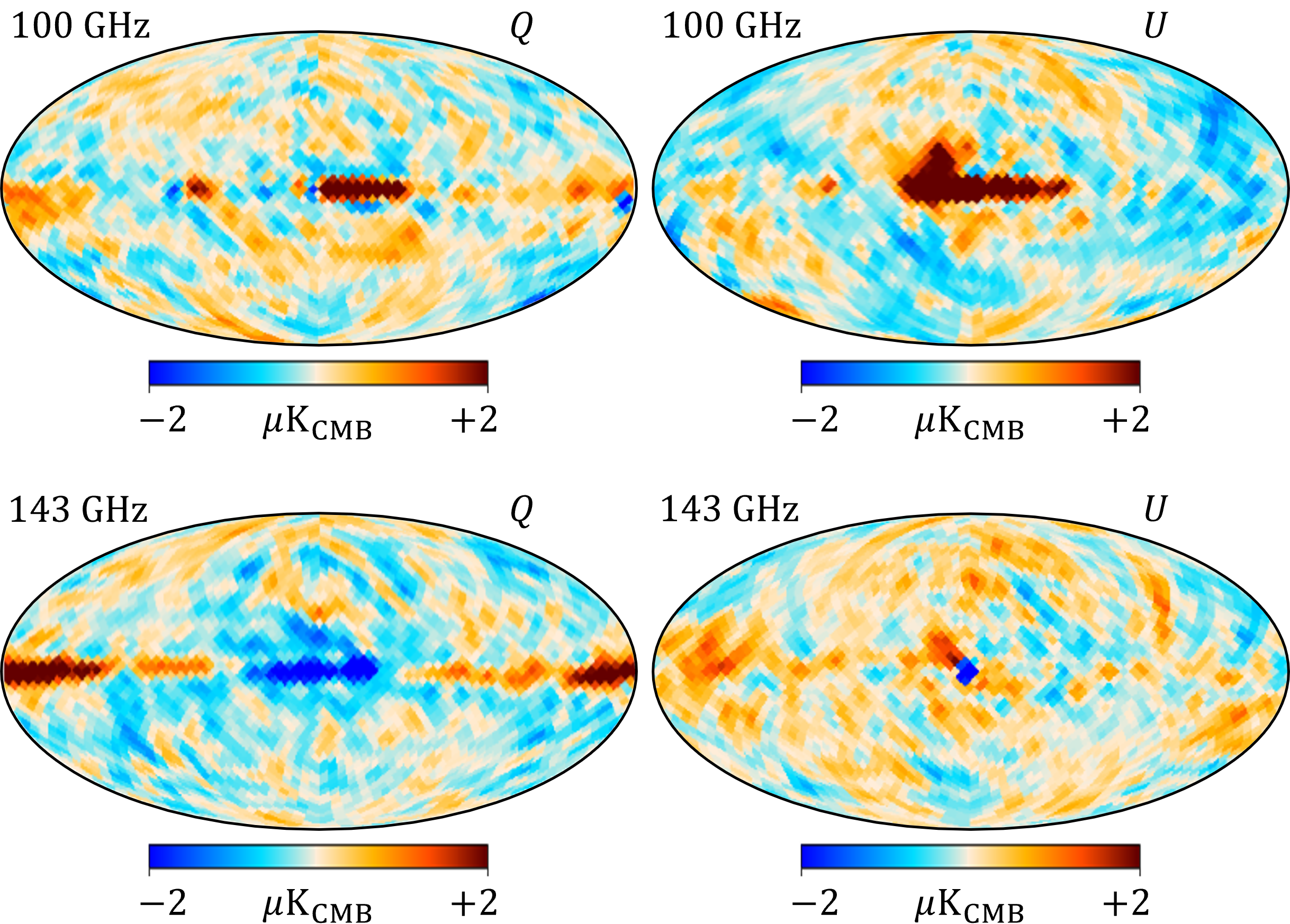}
    \caption{\srolltwo{} data products of the \planck{} $Q$ and $U$ maps at frequencies $100$ GHz and $143$ GHz, after component separation, used in this work, displayed in Galactic coordinates.\label{fig:planck}}
\end{figure}

The goal of this work is the analysis of the \srolltwo{} \planck{} polarization data products \citep{Delouis:2019bub}. They consist of Stokes $Q$ and $U$ maps at the $100$ GHz and $143$ GHz HFI frequency channels, stored at pixel resolution $\nside=16$. The \planck{} maps are first smoothed with cosine beam window functions, and then cleaned from foreground contamination through a template fitting procedure \citep{2020A&A...635A..99P}. Figure~\ref{fig:planck} shows the map products in Galactic coordinates. We note that close to the Galactic plane, $Q$ and $U$ on both channels are visibly contaminated by residual systematic effects, which we masked prior to the analysis in order to avoid bias. The arc-shaped features in the northern and southern Galactic hemisphere likely indicate residual gain variations caused by the ADCNL systematic effect. As shown by \cite{Delouis:2019bub}, these features show lower residual power than the CMB in the 100$\times$143~GHz $EE$ cross-spectrum but may still amount to a nonnegligible bias in cosmological analyses.

\subsection{Masks}

\begin{figure}
    \centering
    \includegraphics[width=\columnwidth]{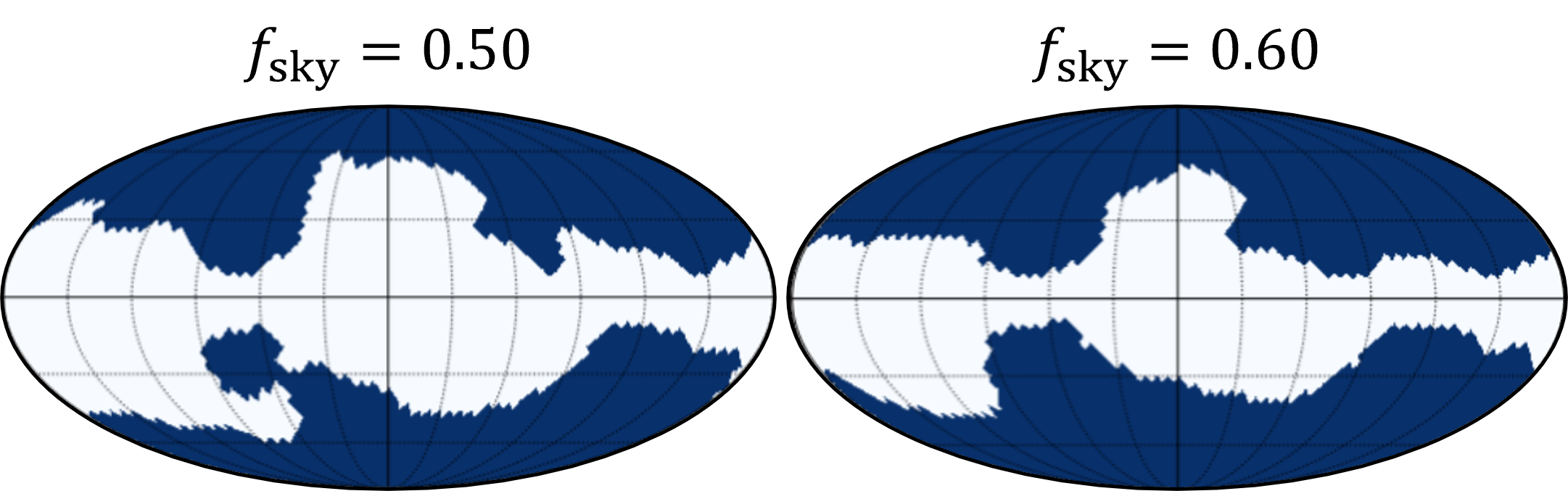}
    \caption{Smoothed version of the \srolltwo{} sky masks at sky fractions $50\%$ and $60\%$ used in this paper, displayed in Galactic coordinates. \label{fig:masks}}
\end{figure}

At low Galactic latitudes, the Milky Way emits polarized foreground radiation which dominates the CMB signal in intensity and polarization. Even after component separation, residuals of this emission need to be excluded from the analysis to avoid biasing cosmological analyses. We therefore applied masks to all maps described in the previous sections. We considered two of the binary polarization masks published in \cite{2020A&A...635A..99P}, retaining sky fractions of $f_{\rm sky}=\{50\%, 60\%\}$. We smoothed them with Gaussian beams of corresponding FWHM of $\{15^\circ, 16^\circ\}$, and apply a binary threshold, setting all pixels with a value larger than 0.5 to one and all others to zero. This procedure allows us to avoid fuzzy borders and mitigate groups of isolated masked pixels. The smoothed masks are shown in Fig.~\ref{fig:masks}. Our baseline mask in this paper is the $f_{\rm sky}=0.5$ smoothed mask, as it retains enough large-scale information to constrain $\tau$ but avoids excessive levels of foregrounds in the Galactic plane.

\section{NN inference}\label{sec:nn_inference}

In this work, we use CNNs to build simulation-based empirical models to perform cosmological inference. In the following, we describe our CNN implementation and give details on the procedures applied to train and test our model on simulations.

\subsection{CNN architecture for $\tau$ estimation}
    
CNNs are the industry standard of pattern recognition in two-dimensional images, performing both classification (e.g., identifying families of objects) and regression tasks (e.g., estimating continuous parameters on maps). The success of CNNs in extracting low-dimensional information from visual input is due to a multilayer image filtering algorithm. This typically involves searching for distinct sets of local features in the original image (through convolution) and compressing the data (through so-called pooling layers), going to lower and lower resolution, before inferring the desired summary statistic.

In our case, we want to retrieve information from data projected on the sphere, requiring convolutions on the spherical domain. To this end, we made use of the \nnhealpix{}\footnote{\url{https://github.com/ai4cmb/NNhealpix}} algorithm which allows to build deep spherical CNNs taking advantage of the \healpix{} tessellation. In particular, \nnhealpix{} performs convolution by looking at the first neighbors for each pixel on the map, and average pooling by downgrading the map resolution (i.e., by going to lower $\nside$ parameter). We refer to \cite{2019A&A...628A.129K} for additional details on how the algorithm works, as well as its advantages and disadvantages. In this work, we used \nnhealpix{} in combination with the public {\tt keras} python package\footnote{\url{https://keras.io}} to build our deep CNN architecture, and to perform training, validation, and testing.

\begin{figure}
\centering
\includegraphics[width=\columnwidth]{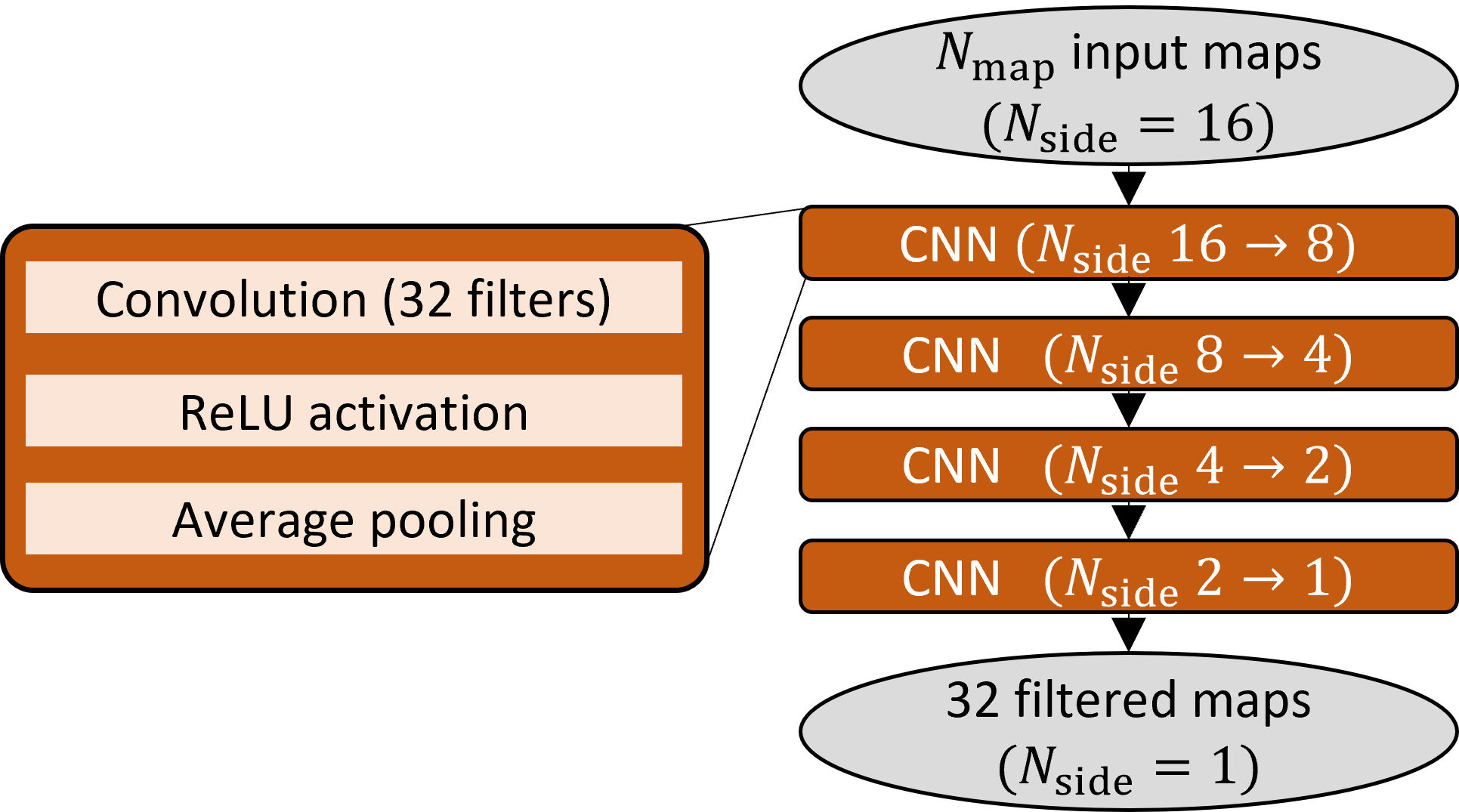}
\caption{Schematic of the convolutional layers of the neural network used in this paper. This represents the first part of the architecture, performing image filtering. \label{fig:cnn}}
\end{figure}

The first part of our CNN, performing image filtering, consists of four CNN building blocks, as illustrated in Fig.~\ref{fig:cnn}. We accept $N_{\rm map}$ input maps, which in our case represent one or two frequency channels and Stokes $Q$ and $U$ maps, hence $N_{\rm map}=2$ or 4. Each convolutional layer introduces 32 filters with nine trainable pixel weights, respectively, and is followed by a Rectified Linear Unit ({\tt ReLU}) activation layer. Mathematically, this means each image pixel $p_i$ undergoes a linear transformation $f$ followed by a nonlinear transformation $g$
    \begin{align}
        & p_i \mapsto p_i'= (f\circ g)(p_i)\, , \\
        & f(p_i) = p_i w_0 + \sum_{j=1}^{N_{\rm neigh}(i)} p_{k_j(i)} w_j\, ,\\
        & g(x)\equiv \max(0,x)\; , \label{eq:relu}
    \end{align}
where $k_j(i)$, $j=1,\dots,\, N_{\rm neigh}(i)$ runs over the indices of all neighboring pixels in the \healpix{} map (which can be either seven or eight, depending on the pixel location). Then, an ``average pooling'' degradation layer reduces the map resolution from $\nside$ to $\nside/2$, assigning to every low-resolution pixel the average of its four children at the next higher resolution. Up to this point, the application of the four CNN building blocks transform the array of input maps at $\nside=16$ (or $N_{\rm pix}=3072$ pixels) into an array of 32 filtered maps at $\nside=1$ (or $N_{\rm pix}=12$ pixels). This represents the image filtering part, meaning the transformation of the original inputs into 32 maximally compressed feature maps that, ideally, retain all the desired (cosmological) information. We still need to ``learn'' the mapping from theses feature maps to the output numbers $\tau_{\rm NN}$ and $\sigma_{\rm NN}(\tau)$ described in the following section. Compression is achieved by two fully connected (or dense) layers. \\

    \begin{figure}
    \centering
    \includegraphics[width=\columnwidth]{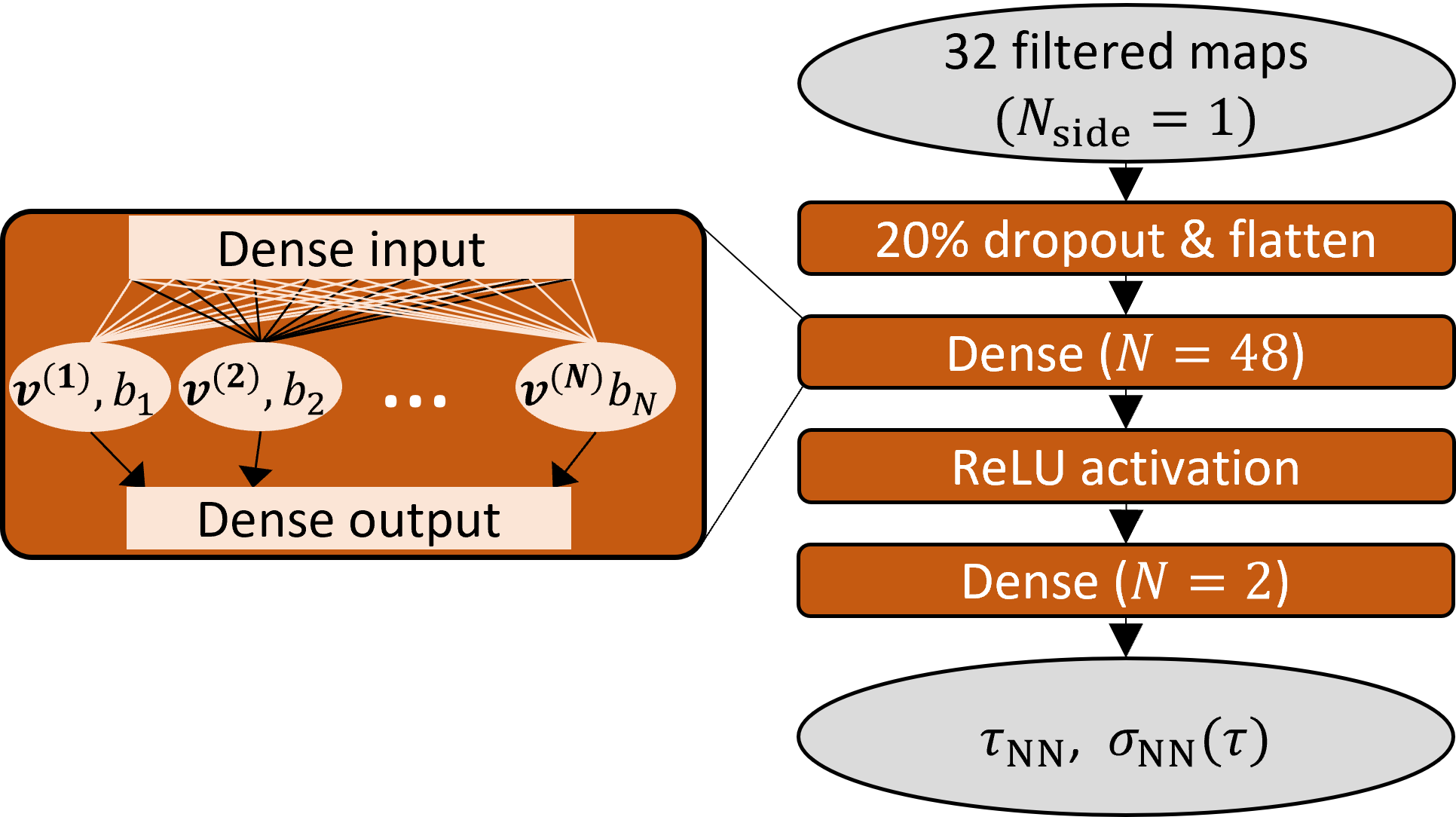}
    \caption{Schematic of the fully connected layers of the neural network used in this paper. This represents the second part of the architecture, performing data compression. }\label{fig:dense}
    \end{figure}
    
A fully connected layer is a linear map from $M$-dimensional input feature space to $N$-dimensional output feature space and is commonly used for data compression (in which case $N<M$). A fully connected layer of output dimension $N$ is said to contain $N$ neurons associated to a vector of trainable weights that parameterize the layer. In each of its $N$ neurons, a fully connected layer linearly contracts the input vector $x$ of length $M$ to a number by means of a weights vector $v^{(i)}$, 
\begin{equation}
    x_i \mapsto x'_{i'} = \sum_{j=1}^M v_j^{(i')} x_j\; .    
\end{equation}
The second part of our CNN, the data compression block, is shown in Fig.~\ref{fig:dense} and contains a dropout and flattening layer, a fully connected layer with 48 neurons, a {\tt ReLU} nonlinear activation layer, concluded by a final fully connected layer with two neurons that outputs $\tau_{\rm NN}$ and $\sigma_{\rm NN}(\tau)$ as described in the following section. The dropout layer acts as a selective off switch for parts of the following fully connected layer, deactivating at random 20$\%$ of its 48 neurons at a time, thus mitigating the overfitting problem common for neural networks \citep{JMLR:v15:srivastava14a}. With the described architecture the total number of weights that need to be optimized during training is $N_{\rm w}\approx4.7\times10^4$.
    
\subsection{Training}\label{ssec:training}

When we train a neural network, we effectively tune its many free parameters until the task at hand, such as estimating parameters from maps, would be optimally performed on the training data. In the following, we describe this procedure in detail.

At each training step we passed one batch of $N_{\rm batch}=32$ training simulations through the network, meaning we simultaneously considered the results from all simulations that belong to a single batch. Input maps need to be masked with the same mask that is used in the analysis. The output values of the two neurons of the final layer, representing the estimated parameters $\tau_{{\rm NN}j}$, $\sigma_{\rm NN}(\tau)_j$ ($j=1,\,\dots,\, N_{\rm batch}$), as well as the truth values $\tau_j$, are then inserted into the loss function \citep{Jeffrey:2020itg}
\begin{align}
    & \mathcal{L}\left[ \tau,\; \left( \tau_{{\rm NN}j},\, \sigma_{\rm NN}(\tau)_j \right) \right] = \notag\\
    & \sum_{j=1}^{N_{\rm batch}} \left[ (\tau_j - \tau_{{\rm NN}j})^2 + \left( (\tau_j - \tau_{{\rm NN}j})^2 - \sigma_{\rm NN}(\tau)_j^2 \right)^2 \right] \; . \label{eq:loss_function}
\end{align}

We then updated all $N_{\rm w}$ network parameters subject to minimizing this loss function. For doing so, we used the {\sc Adam} optimizer, a widely used stochastic gradient descent algorithm implemented in {\tt keras}, for which we found an initial training rate of ${\sc LR}=10^{-3}$ and first- and second-moment exponential decay rates $\beta_1=0.9$ and $\beta_2=0.999$ to be appropriate. Repeating the described procedure for the entire training set of size $N_{\rm train}=190,000$ made up one training epoch\footnote{Among the total 200,000 simulations generated as described in Sect.~\ref{sec:simulations_data}, we actually used 190,000 to optimize the NN's parameters, while we used the remaining 10,000 as a validation set. }. We trained on a maximum of 45 epochs, using the {\tt keras} callback function {\tt ReduceLROnPlateau} to allow for learning rates to decrease by a factor of 0.1 if the loss of the validation set did not improve over the course of five epochs. Moreover, the callback function {\tt EarlyStopping} allows for training to stop after a minimum number of epochs (in our case 20) without improvement in the validation loss. Using both of these callback functions allowed for a faster convergence and suppressed unwanted oscillations in the loss function during the training phase. Training on a 32-core Intel Xeon CPU node took about three hours, while training on eight NVIDIA Tesla V100 GPU cores took about 30 minutes.

\subsection{Testing}

After training, we fix the neural network parameters, which in principle completes the model building. However, trained NNs may not perform well for two main reasons: firstly, the loss function may have not converged to its global minimum, causing model predictions to be biased. Secondly, the model may overfit the input, meaning that the network learns the training set's features with an excellent accuracy, but fails to make correct predictions on similar, independent test sets. One usually addresses both problems by testing the model's predictions on simulations that have not been fed into the network before. We used 2$\times$3 test sets of 10,000 sky simulations with fixed input $\tau=\{0.05$, 0.06, 0.07$\}$, described in detail in Sects.~\ref{ssec:noise_sims} and \ref{ssec:sroll2_sims}. 

We note that, by inferring only $\tau_{{\rm NN}}$ and $\sigma_{\rm NN}(\tau)$, we implicitly assumed Gaussian posteriors, which we exhaustively validated on simulations by checking for biases in the Gaussian mean and variance (see Sect.~\ref{sec:results_sims}). If, instead, our algorithm had provided an entire, potentially non-Gaussian probability distribution function or higher statistical moments, we would have needed to perform more extensive sanity checks and indicate credible intervals instead of Gaussian standard deviations.

\section{Results on simulations}
\label{sec:results_sims}
Before arriving at the estimation of $\tau$ from the \planck{} \srolltwo{} data, we considered several setups to train our CNN model, increasing the complexity of the training simulations. This allowed us to get valuable insight into the learning process. In particular, we started by training the CNN on a set of simulations including CMB with Gaussian noise (see Sect.~\ref{ssec:noise_sims}), either on a single frequency channel, or on two channels. We then moved to simulations including non-Gaussian systematic effects (i.e., \srolltwo{} simulations), trying different possible strategies to obtain unbiased $\tau$ estimates in the presence of complex residuals. Only once we achieved this, we applied our trained model to real \planck{} data. In all the cases presented in this section, we trained and tested the CNNs considering the $f_{\rm sky}=0.5$ mask as our reference (see Fig.~\ref{fig:masks}). A summary of all analysis cases, along with their corresponding results tables and figures, can be found in Table~\ref{tab:tests_overview}.

\begin{table*}
\centering
\caption{References to results tables and figures in this paper.}\label{tab:tests_overview}
    \begin{tabular}{lccc}
    \hline\hline
    \noalign{\smallskip}
     &  Gaussian test simulations & \srolltwo{} test simulations & \planck{} data \\
     \noalign{\smallskip}
    \hline
    \noalign{\smallskip}
    Gaussian NN (one channel) & Table~\ref{tab:gaussian_training}; Fig.~\ref{fig:hist_1ch_vs_2ch} & Table~\ref{tab:gauss_training_sr_test}; Fig.~\ref{fig:hist_1ch_vs_2ch} & \\
    Gaussian NN (two channels) & Table~\ref{tab:gaussian_training}; Figs.~\ref{fig:hist_1ch_vs_2ch},~\ref{fig:hist_3x2} & Table~\ref{tab:gauss_training_sr_test}; Figs.~\ref{fig:hist_1ch_vs_2ch},~\ref{fig:hist_3x2} & Table~\ref{tab:planck_results_masks}; Figs.~\ref{fig:planck_nn_variation},~\ref{fig:planck_masks}  \\
    HL likelihood & Table~\ref{tab:gaussian_training} & Table~\ref{tab:gauss_training_sr_test} & \\
    \srolltwo{} training & & Table~\ref{tab:sroll2_training} & \\
    \srolltwo{} retraining & & Table~\ref{tab:sroll2_training}; Fig.~\ref{fig:hist_3x2} & Table~\ref{tab:planck_results_masks}; Figs.~\ref{fig:planck_nn_variation},~\ref{fig:planck_masks} \\
    Empirical likelihood & & & Table~\ref{tab:planck_results_masks}; Fig.~\ref{fig:planck_masks} \\
    \hline\hline
    \end{tabular}
\end{table*}

\subsection{Gaussian training}\label{sec:gaussian_training}
 
As aforementioned, we first tested the ability of our CNN to estimate the value of $\tau$ considering only Gaussian noise. These simulations have noise amplitudes and  pixel-pixel correlations directly estimated from \planck{} maps, and therefore serve as a good description of the Gaussian noise present in real data. At the same time, they lack for realism, since they do not include non-Gaussian residual systematic effects, contamination due to Galactic foregrounds, both known to be present on the \planck{} \srolltwo{} maps. We therefore expected these models (which we refer to as ``Gaussian models'') to induce a bias on $\tau$ when applied to the more realistic \srolltwo{} simulations, or to real \planck{} data.

\subsubsection{Single channel}

We began by training our CNN on Stokes $Q$ and $U$ maps with Gaussian \planck-like noise and CMB at $143$ GHz only, thus feeding $N_{\rm map}=2$ maps to the network. In the left-hand side of Table~\ref{tab:gaussian_training}, we show the results of testing $N_{\rm sims}=10,000$ Gaussian simulations of CMB and noise generated with fiducial $\tau=0.05$, $0.06$, and $0.07$, respectively. The average learned mean posterior values $\overline{\tau_{\rm NN}}$ are close to unbiased and deviate at the 0.2$\sigma$ level. The average learned posterior standard deviations $\overline{\sigma_{\rm NN}(\tau)}$ are within $5\%$ agreement with the sample scatter across simulations, $\sigma(\tau_{\rm NN})$.

To assess the performance of the Gaussian model also on non-Gaussian \planck-like maps, we tested this model on 10,000 \srolltwo{} simulations generated with fiducial $\tau=0.06$ (see Sect.~\ref{ssec:sroll2_sims}). As illustrated in the upper right panel of Fig.~\ref{fig:hist_1ch_vs_2ch}, this leads to a bias of more than $1\sigma$ on $\tau_{\rm NN}$. These tests on a single frequency channel leave us with two conclusions: on the one hand, CNNs are able to correctly retrieve $\tau$ and its statistical uncertainty from a single \planck-like simulation of the 143~GHz channel containing correlated Gaussian noise. On the other hand, systematic effects present in the \planck{} \srolltwo{} simulations bias the single-channel CNN inference, as expected. 
To improve our results, we added another frequency channel to the inference pipeline, seeking to mitigate this bias. We expected that combining two channels should lead to a lower error bar and a lower bias on the \srolltwo{} simulations, in a similar way as cross-spectra achieve lower noise bias than auto-spectra.
    
\subsubsection{Two channels}
    
\begin{figure}
        \centering
        \includegraphics[width=\columnwidth]{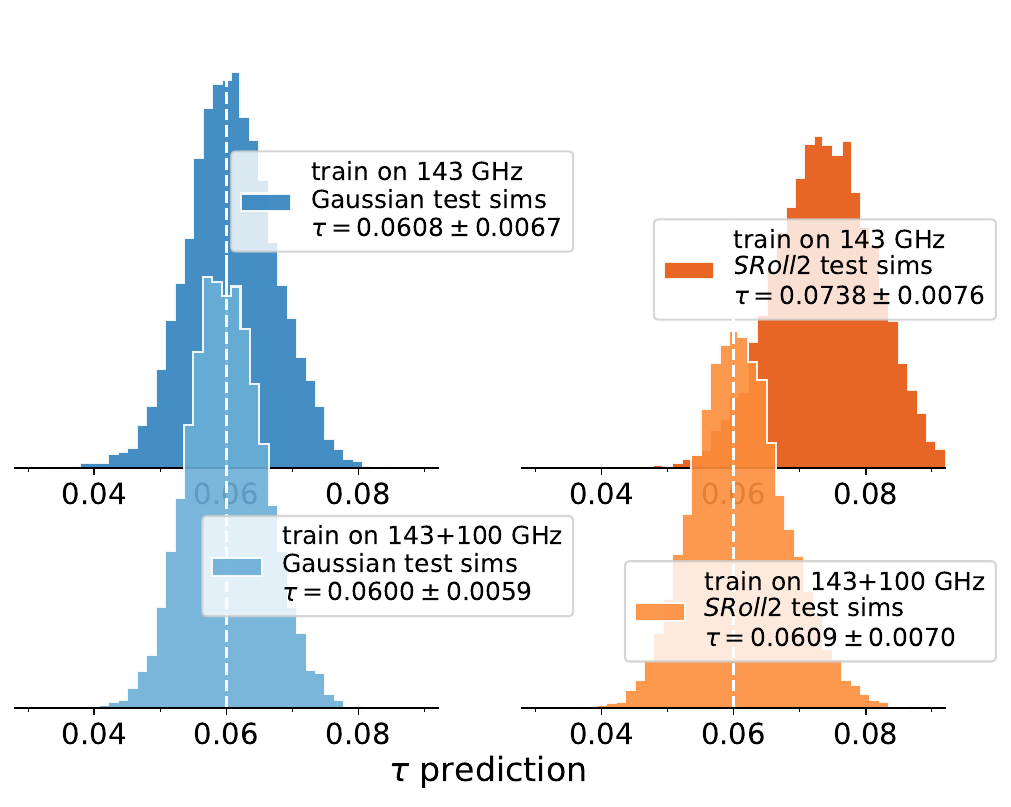}
        \caption{Predictions of $\tau_{\rm NN}$ from $10,000$ simulations with input $\tau=0.06$, containing either CMB with Gaussian noise (\emph{left panels}) or CMB with \srolltwo{} noise + systematics  (\emph{right panels}). The two rows denote different CNN models trained on CMB with Gaussian noise on a single frequency channel (\emph{top}), on two frequency channels (\emph{bottom}). \label{fig:hist_1ch_vs_2ch}}
    \end{figure}

As a second test, we added the HFI 100~GHz channel in the training and testing procedures, simulated as CMB plus the corresponding Gaussian correlated noise, so that $N_{\rm map}=4$ maps were fed into the neural network. The results from testing on Gaussian noise are shown in Table~\ref{tab:gaussian_training}. We note two positive effects: firstly, the small bias observed for Gaussian noise on a single channel reduces to below $1\%$ of a standard deviation. Secondly, the learned $\sigma_{\rm NN}(\tau)$ decreases by more than $10\%$ when training on two frequency channels. Meanwhile, the prediction of the posterior standard deviation stays within $5\%$ of the sample standard deviation of the inferred $\tau_{\rm NN}$. The same results are presented in Fig.~\ref{fig:hist_1ch_vs_2ch} for fiducial $\tau=0.06$, showing significant improvement of the two-channel CNN inference in the lower panels with respect to the one-channel results (upper panels). We proceeded to test this two-channel Gaussian model on the \srolltwo{} simulations. As shown in the right panel of Fig.~\ref{fig:hist_1ch_vs_2ch}, for fiducial $\tau=0.06$, the addition of a second channel allows for a significant reduction of the systematic-related bias in $\tau_{\rm NN}$ and to a better statistical constraint. This led us to conclude that CNNs are able to recognize common features across channels, combining the information to reduce the statistical uncertainty and the bias due to uncorrelated systematic effects.

The corresponding quantitative results, for all the three $\tau$ values used during testing, are listed in Table~\ref{tab:gauss_training_sr_test}: adding a second channel in the Gaussian training model leads to improved results on the \srolltwo{} test simulations for all considered values of $\tau$. However, a residual bias is still present, especially for $\tau=0.05$, when the CMB signal is smallest.

Moreover, we noticed that, when applied to the \srolltwo{} test maps, the models trained on Gaussian simulations returned values of $\sigma_{\rm NN}(\tau)$ that disagreed with the actual spread of estimates $\sigma(\tau_{\rm NN}$), with the latter being up to $\sim 25\%$ larger. This implies that the learned error was not accurate in this case, hence could not be used to describe the uncertainties of our inferred $\tau$ values on \srolltwo{} maps. We address this issue in Sect.~\ref{ssec:error_bars}.

\begin{table*}
\centering
\caption{$\tau$ predictions from $10,000$ Gaussian CMB + noise simulations generated with three different, fixed fiducial $\tau$ values. The results correspond to the Gaussian NN training on one and two channels, and the Bayesian inference with a power spectrum likelihood. We show the posterior mean $\tau_{\rm NN/HL}$ and standard deviation $\sigma_{\rm NN/HL}(\tau)$ averaged over all simulations, as well as the scatter of $\tau_{\rm NN/HL}$ over all simulations.}\label{tab:gaussian_training}
    \begin{tabular}{cccccccccc}
    \hline\hline
    \noalign{\smallskip}
    \multicolumn{10}{c}{Test on Gaussian simulations}\\
     \noalign{\smallskip}
    \hline
    \noalign{\smallskip}
     & \multicolumn{3}{c}{143~GHz} & \multicolumn{3}{c}{143+100~GHz} & \multicolumn{3}{c}{143$\times$100~GHz}\\
    \noalign{\smallskip}

     & \multicolumn{3}{c}{Gaussian training} & \multicolumn{3}{c}{Gaussian training} & \multicolumn{3}{c}{HL likelihood}\\
    \hline
    \noalign{\smallskip}

    fiducial $\tau$ & $\overline{\tau_{\rm NN}}$ & $\overline{\sigma_{\rm NN}(\tau)}$ & $\sigma(\tau_{\rm NN})$ & $\overline{\tau_{\rm NN}}$ & $\overline{\sigma_{\rm NN}(\tau)}$ & $\sigma(\tau_{\rm NN})$ & $\overline{\tau_{\rm HL}}$ & $\overline{\sigma_{\rm HL}(\tau)}$ & $\sigma(\tau_{\rm HL})$ \\
    \noalign{\smallskip}
    0.05 & 0.0508 & 0.0059 & 0.0066& 0.0503 & 0.0054 & 0.0057 & 0.0496 & 0.0046 & 0.0047 \\ 
    0.06 & 0.0608 & 0.0065 & 0.0067& 0.0600 & 0.0056 & 0.0059 & 0.0596 & 0.0048 & 0.0048 \\ 
    0.07 & 0.0712 & 0.0067 & 0.0070& 0.0702 & 0.0057 & 0.0063 & 0.0697 & 0.0048 & 0.0049 \\ 
      \noalign{\smallskip}
    \hline\hline
    \end{tabular}
\end{table*}

\begin{table*}
\centering
\caption{Same as Table~\ref{tab:gaussian_training} but testing on CMB and \srolltwo{} simulations instead of CMB and Gaussian noise simulations.}
\label{tab:gauss_training_sr_test}
    \begin{tabular}{cccccccccc}
    \hline\hline
    \noalign{\smallskip}
    \multicolumn{10}{c}{Test on \srolltwo{} simulations}\\
     \noalign{\smallskip}
    \hline
    \noalign{\smallskip}
     & \multicolumn{3}{c}{143~GHz} & \multicolumn{3}{c}{143+100~GHz} & \multicolumn{3}{c}{143$\times$100~GHz}\\
    \noalign{\smallskip}

     & \multicolumn{3}{c}{Gaussian training} & \multicolumn{3}{c}{Gaussian training} & \multicolumn{3}{c}{HL likelihood}\\
    \hline
    \noalign{\smallskip}

    fiducial $\tau$ & $\overline{\tau_{\rm NN}}$ & $\overline{\sigma_{\rm NN}(\tau)}$ & $\sigma(\tau_{\rm NN})$ & $\overline{\tau_{\rm NN}}$ & $\overline{\sigma_{\rm NN}(\tau)}$ & $\sigma(\tau_{\rm NN})$ & $\overline{\tau_{\rm HL}}$ & $\overline{\sigma_{\rm HL}(\tau)}$ & $\sigma(\tau_{\rm HL})$ \\
    \noalign{\smallskip}
    0.05 & 0.0669 & 0.0065 & 0.0074& 0.0536 & 0.0055 & 0.0067 & 0.0478 & 0.0050 & 0.0079 \\ 
    0.06 & 0.0738 & 0.0067 & 0.0076& 0.0609 & 0.0056 & 0.0070 & 0.0585 & 0.0050 & 0.0073   \\ 
    0.07 & 0.0813 & 0.0069 & 0.0074& 0.0690 & 0.0057 & 0.0071 & 0.0688 & 0.0049 & 0.0069  \\ 
    \hline\hline
    \end{tabular}
\end{table*}

\subsection{Comparison with Bayesian inference from cross-QML power spectrum estimates}\label{ssec:likelihood}

In this section we compare NN inference results with results coming from a standard Bayesian approach applied to $E$-mode power spectra. In particular, we considered quadratic Maximum Likelihood (QML) estimates \citep[see, e.g.,][]{Tegmark:2001zv} of the 100$\times$143~GHz $EE$ cross-spectrum and drew posterior samples using the well-known power spectrum likelihood approximation introduced by \cite{hamimechelewis2008} (in the following HL likelihood). The HL likelihood provides a good approximation to the non-Gaussian distribution of the exact power spectrum likelihood, which markedly differs from Gaussianity at the low multipoles $2\leq\ell\lesssim30$ that are most relevant for constraining $\tau$. Evaluating the HL likelihood requires a power spectrum covariance matrix, which we obtained directly from simulations of Gaussian noise and CMB realized with the same $\tau$ values used for generating the test simulations (Sect.~\ref{sec:simulations_data}). For the HL likelihood we assumed a theoretical model of the CMB $E$-modes, computed with {\sc CAMB}, considering the multipole range $2\leq\ell\leq30$, and sampling only for the $\tau$ parameter, keeping $10^9 A_{s} e^{-2\tau} = 1.884$ fixed. Our final results are the best-fit value $\tau_{\rm HL}$, the standard deviation $\sigma_{\rm HL}(\tau)$ of the posterior, and the scatter $\sigma(\tau_{\rm HL})$ computed from the set of test simulations.

We ran the HL likelihood on $3\times10,000$ Gaussian sky simulations with input  $\tau=0.05,\,0.06$, and 0.07. As shown in the last three columns of Table~\ref{tab:gaussian_training}, we find unbiased best-fit results with average posterior standard deviation $\overline{\sigma_{\rm HL}(\tau)}$ and best-fit parameter scatter $\sigma(\tau_{\rm HL})$ of $\sim 0.0048$. We note that the uncertainties derived from sampling the HL likelihood are $\sim20\%$ smaller than the ones from NN estimates. Part of the scatter of $\tau_{\rm NN}$ comes from the intrinsic stochastic nature of the training process. We could reduce this scatter by taking the average over multiple NN models (as discussed in Sect.~\ref{ssec:error_bars}). Nevertheless, these results reveal that although we were able to retrieve unbiased $\tau$ values with NNs from Gaussian simulations, our estimator does not achieve minimum variance. Further development of the method, including an optimization of the convolution algorithm on the sphere, the NN architecture, and the training procedure, are required and will be explored in future work in the light of improving the estimator's variance.

\begin{figure}
    \centering
    \includegraphics[width=\columnwidth]{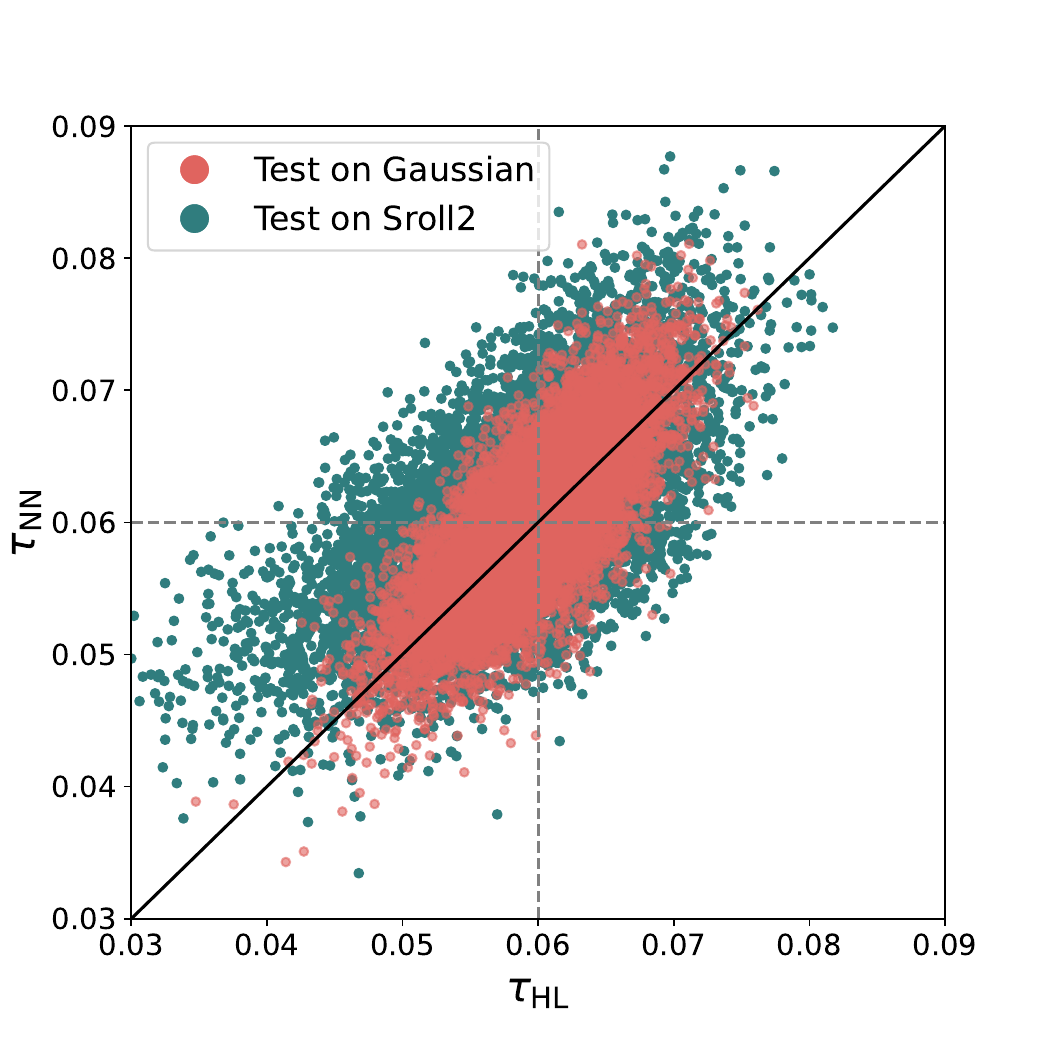}
    \caption{Per-simulation comparison between the HL likelihood estimate $\tau$ and the NN estimate $\tau_{\rm NN}$ for a test set of 10,000 simulations realized with $\tau=0.06$. Gaussian simulations are shown in bright red, \srolltwo{} simulations in dark green. The correlation coefficients between both estimators are $76\%$ (Gaussian) and $63\%$ (\srolltwo).}.  \label{fig:NN_HL_comparison}
\end{figure}

In addition to Gaussian simulations, we applied the cross-spectrum inference pipeline on $3\times10,000$ \srolltwo{} simulations and show the corresponding results in the last three columns of Table~\ref{tab:gauss_training_sr_test}. We stress that the HL likelihood contains the same covariance matrix as before, calculated from Gaussian simulations. This is done in analogy with the case of Gaussian NN training applied to \srolltwo{} simulations, therefore neglecting the presence of systematic effects. We retrieve biased estimates on $\tau$, confirming our expectation that the power spectrum model implemented in the likelihood is an inaccurate representation of the \srolltwo{} simulations, which include spurious non-Gaussian signals. Interestingly, this affects the NN and HL estimates in different ways, leading to biases in opposite directions for $\tau=0.05$ and $0.06$. To study the relative behavior of the two estimators, it is instructive to look at a one-by-one comparison of the NN and HL results on the same 10,000 test simulations, as presented in Fig.~\ref{fig:NN_HL_comparison} for $\tau=0.06$. The scatter plot of the estimated $\tau_{\rm NN}$ and $\tau_{\rm HL}$ on Gaussian simulations and on \srolltwo{} simulations are shown in bright red and dark green, respectively. In the Gaussian case the correlation of the estimated $\tau$ values is at a level of $\sim76\%$, while for \srolltwo{} it is at $63\%$.
We conclude that map-level systematic effects, which are partially unaccounted for in the estimates, decrease the correlation and increase the differences between $\tau_{\rm HL}$ and $\tau_{\rm NN}$ when changing from Gaussian to \srolltwo{} test simulations. This indicates that spurious non-Gaussian signals impact the two estimators in different ways.

\subsection{Training including systematic effects}

As previously seen, the two-channel Gaussian training allowed to improve our $\tau$ estimates on \srolltwo{} simulations. However, the continued occurrence of bias, even though small, motivated us to evolve the training setup by including systematic effects in the training set. Our goal was to achieve fully unbiased results before applying our NN models to real \planck{} maps. In this section we explore two possible ways of including systematics: training on \srolltwo{} simulations from the very beginning and performing a \srolltwo{} retraining update on previously trained Gaussian networks. 

\subsubsection{Training on \srolltwo{} simulations}
\label{Sec:trainingSR}
\begin{table*}
\centering
\caption{$\tau$ predictions from $10,000$ CMB + \srolltwo{} test simulations generated with three different fiducial $\tau$ values. These results correspond to two frequency channels, either training on \srolltwo{} from the start or retraining on \srolltwo{} maps. Displayed are the average posterior mean, average predicted standard deviation $\overline{\sigma_{\rm NN}(\tau)}$, and the scatter $\sigma(\tau_{\rm NN})$ calculated across the test simulations. }\label{tab:sroll2_training}
    \begin{tabular}{ccccccc}
    \hline\hline
     \noalign{\smallskip}
    \multicolumn{7}{c}{Test on \srolltwo{} simulations}\\
     \noalign{\smallskip}
    \hline
    \noalign{\smallskip}
     &  \multicolumn{3}{c}{143+100~GHz} & \multicolumn{3}{c}{143+100~GHz}\\
     &  \multicolumn{3}{c}{\srolltwo{} training} & \multicolumn{3}{c}{\srolltwo{} retraining}\\
     \noalign{\smallskip}
    \hline
    \noalign{\smallskip}
    fiducial $\tau$ & $\overline{\tau_{\rm NN}}$ & $\overline{\sigma_{\rm NN}(\tau)}$ & $\sigma(\tau_{\rm NN})$ & $\overline{\tau_{\rm NN}}$ & $\overline{\sigma_{\rm NN}(\tau)}$ & $\sigma(\tau_{\rm NN})$ \\
    \noalign{\smallskip}
    0.05 & 0.0526 & 0.0059 & 0.0066& 0.0508 & 0.0077 & 0.0091 \\ 
    0.06 & 0.0622 & 0.0062 & 0.0070& 0.0606 & 0.0079 & 0.0088 \\ 
    0.07 & 0.0722 & 0.0064 & 0.0070& 0.0707 & 0.0081 & 0.0087 \\ 
    \hline\hline
    \end{tabular}
\end{table*}

The \srolltwo{} simulations \citep{Delouis:2019bub} are designed to accurately describe \planck's Gaussian noise component and non-Gaussian polarization systematics. Motivated by this, we trained a CNN from the start on the 200,000 \srolltwo{} training simulations described in Sect.~\ref{ssec:sroll2_sims}. As usual, we used 190,000 simulations to perform weight optimization, and 10,000 for validation. We trained on \planck's 143~GHz and 100~GHz channels simultaneously and used the same hyperparameter values as for the Gaussian training, described in Sect.~\ref{ssec:training}. We stress that even though artificially augmented by forming new channel pair combinations, the \srolltwo{} training set was essentially built from $400$ sampled skies only. We tested on $3\times10,000$ \srolltwo{} simulations with fixed $\tau=0.05$, 0.06, and 0.07, generated from the remaining 100 independent realizations that the CNN did not ``see'' during training.

Table~\ref{tab:sroll2_training} shows the results obtained with this approach. For the three input $\tau$ values we find a positive bias of $\sim0.4\sigma$. For $\tau=0.06$, the average learned error $\overline{\sigma_{\rm NN}(\tau)}=0.0062$ is slightly larger than for the two-channel Gaussian training but smaller than the scatter $\sigma(\tau_{\rm NN})=0.0070$. We see similar results both for the Gaussian CNN and the HL likelihood inference (see Table~\ref{tab:gauss_training_sr_test}). As in the case of Gaussian NN training, the learned error does not agree with the \srolltwo{} simulation scatter, therefore it cannot be used to infer the statistical uncertainty on $\tau_{\rm NN}$.

\begin{figure}
    \centering
    \includegraphics[width=\columnwidth]{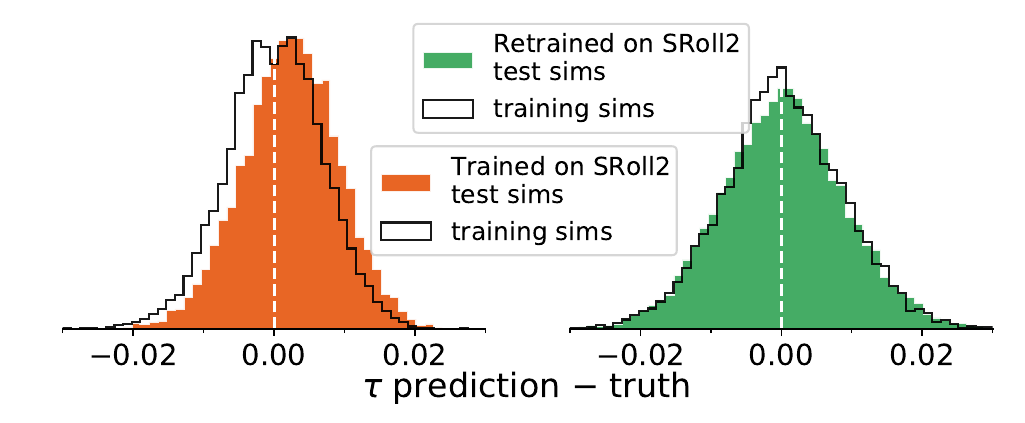}
    \caption{Neural network accuracy in predicting the true $\tau$ input from $10,000$ simulations. Step-filled histograms show the results on unseen test simulations, black outlines show the results on a subset of the actual training simulations. We compare a network exclusively trained on \srolltwo{} simulations (\textit{left panel}) with a Gaussian network retrained on \srolltwo{} simulations (\textit{right panel}).  \label{fig:hist_overfitting}}
\end{figure}

We ascribe the main reason for the bias on $\tau$ to overfitting. Figure~\ref{fig:hist_overfitting} illustrates this problem. We compared the $\tau$ predictions on a set of 10,000 test simulations with the ones coming from 10,000 training simulations. The results show a bias and standard deviation of $\Delta\tau=0.0023\pm0.0069$ for the test set, while the training set is unbiased, with $\Delta\tau=0.0001\pm0.0068$. This is clear evidence for overfitting: while the model performs well on the 400 \srolltwo{} simulations that the training set is built from, these are not enough to generalize to the remaining 100 \srolltwo{} simulations used to build the test set, leading to the observed bias on $\tau$ in the latter case. 

\subsubsection{Retraining update with \srolltwo{} simulations}\label{ssec:retraining}

We recognized the bias described above as a critical problem that needed to be addressed. The obvious option, training on a considerably larger simulation set, was unavailable to us due to the limited number of \srolltwo{} realizations. Therefore, we applied a transfer learning technique to inform our previously trained Gaussian networks on the \srolltwo{} systematics.
As shown in the previous sections, our Gaussian NN model is not affected by overfitting issues and, if trained on two channels, performs reasonably well even on \srolltwo{} simulations. This motivated us to leverage the existing results on Gaussian networks to solve the overfitting issue with as little changes as possible. To this end, we chose the approach of retraining the two-channel Gaussian model on the full set of \srolltwo{} training simulations, while targeting two specific goals:
\begin{itemize}
    \item[(i)] The retrained CNN should learn to extract information on the systematic effects present in the \srolltwo{} simulations and update its CNN weights just enough to achieve fully unbiased results on the \srolltwo{} training set.
    \item[(ii)] At the same time, we wanted to ensure that the information already learned was not destroyed during the new training phase (an issue sometimes referred to as ``catastrophic forgetting'', see e.g., \cite{2017PNAS..114.3521K}, \cite{ramasesh2021effect}), avoiding going back to the overfitting situation described in the previous section.
\end{itemize}
We achieved this by performing what we call ``minimal retraining'': we chose the hyperparameters of the NN such that we obtained unbiased results on the \srolltwo{} test simulations while making minimal changes to the original network. We found an optimal setup with five retraining epochs, a learning rate of $LR=10^{-7}$, and no additional changes to the original network architecture.

    \begin{figure}
        \centering
        \includegraphics[width=\columnwidth]{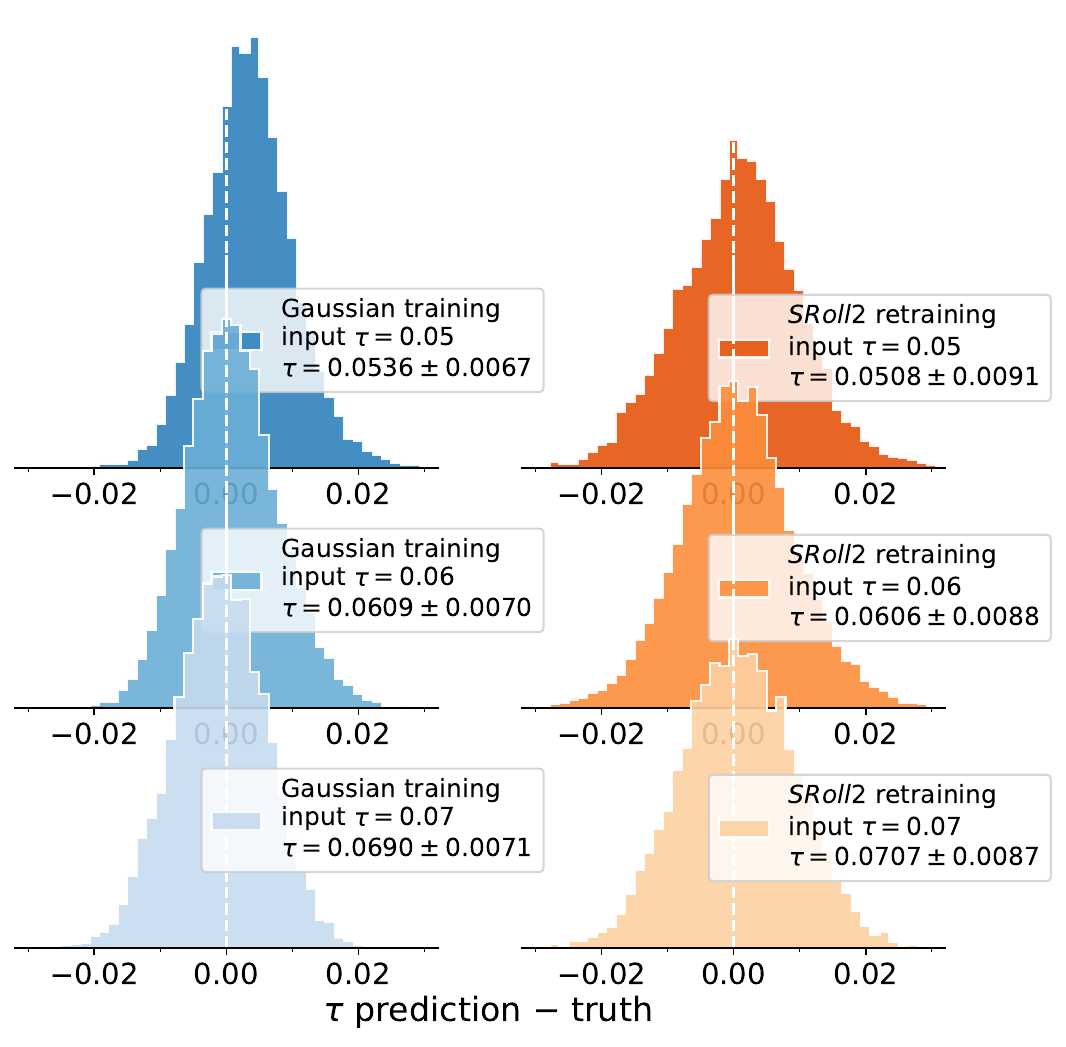}
        \caption{Predictions of $\tau_{\rm NN}$ on $10,000$ \srolltwo{} simulations with input $\tau=0.05$, 0.06, and 0.07 (first, second, third row, respectively). The two columns display two different NN models trained on two channels of Gaussian simulations (left panels) and retrained on two channels of \srolltwo{} simulations (right panels). All results correspond to $f_{\rm sky}=0.5$.  \label{fig:hist_3x2}}
    \end{figure}
    
The right panel of Fig.~\ref{fig:hist_overfitting}, in analogy to the left panel, compares the distribution of $\Delta\tau$ from the \srolltwo-retrained model on training simulations (black contours), or test simulations (green filled histogram). We find both histograms to be in good agreement, indicating that unlike the \srolltwo-trained model, the retrained model does not suffer from overfitting, thus achieving our goal (ii) defined above. 
Table~\ref{tab:sroll2_training} on the right-hand side lists the results of the \srolltwo-retrained model on \srolltwo{} test simulations. We find $\overline{\tau_{\rm NN}}=0.0508$, 0.0606, and 0.0707 for the respective input values of $\tau=0.05$, 0.06, and 0.07. This amounts to a bias below $\Delta\tau=8\times10^{-4}$, or $\lesssim 0.1\sigma$. 
In Fig.~\ref{fig:hist_3x2}, we show a comparison of the results on \srolltwo{} test sets obtained by Gaussian and \srolltwo-retrained CNNs. The reduction of the bias is evident, in particular for $\tau=0.05$. Therefore, we chose the retrained approach as our baseline model to estimate $\tau$ on real \planck{} data. At the same time, this approach brings an increase in $\sigma(\tau_{\rm NN})$, an effect not seen with the \srolltwo{} training procedure described in Sect.~\ref{Sec:trainingSR}\footnote{Compare the fourth column in Table~\ref{tab:sroll2_training} with the seventh column in Table~\ref{tab:gauss_training_sr_test}}. This could be the consequence of the typical variance-bias trade-off observed between statistical estimators: with minimal retraining we are able to achieve unbiased estimates (goal (i) above) at the cost of a larger $\sigma(\tau_{\rm NN})$. In addition to that, we are still unable to retrieve values of the learned $\sigma_{\rm NN}(\tau)$ that agree with $\sigma(\tau_{\rm NN})$ for \srolltwo{} simulations (and therefore also for \planck{} data). We conclude that, except for case in which we test the Gaussian model on Gaussian simulations, we cannot use the learned error as an estimate of the uncertainty of the inferred $\tau_{\rm NN}$.

\subsection{NN errors}\label{ssec:error_bars}

The loss function in Eq. \eqref{eq:loss_function} provides an estimate for the posterior standard deviation $\sigma_{\rm NN}(\tau)$. However, as seen in the previous sections, the learned $\sigma_{\rm NN}(\tau)$ tends to underestimate the actual spread of the inferred values of $\tau_{\rm NN}$ on test set maps, especially in the case of \srolltwo{} maps. We therefore proceeded to empirically estimate our errors from simulations.

In doing so, we needed to make an additional consideration: training a NN is an intrinsically stochastic procedure that relies upon the use of a stochastic optimizer, randomly initialized NN weights and random realizations of the maps in the training set. This results in the fact that each NN prediction can be described as the sum of two random variables: $\tau_{\rm NN} = \tau + \Delta_{\rm NN}$, and therefore

\begin{equation} 
\label{eq:NNerror_model}
    \sigma^2(\tau_{\rm NN}) = \sigma^2(\tau) + \sigma^2(\Delta_{\rm NN})+2\,{\rm Cov}(\tau, \Delta_{\rm NN})\, ,
\end{equation} 

where the first source of uncertainty, $\sigma(\tau)$, is due to noise and cosmic variance of test simulations or observed data, while the second, $\sigma(\Delta_{\rm NN})$, represents the stochasticity of the NN estimator. These two terms are sometime referred to as \emph{aleatory} and \emph{epistemic} error, respectively \citep{nn_err}. 

We can measure the uncertainty related to the NN stochasticity by training an ensemble of models, all based on the same architecture and hyperparameters, but with different initial weights and training set realizations. Our estimate of $\sigma(\Delta_{\rm NN})$ is given by the standard deviation of the models' $\tau$ predictions when tested on a single test map. In practice, we defined the ``model ID'' of a trained NN as the fixed random seed controlling the initialization of network weights. We generated a new training set of simulations whose specific realizations (of CMB, noise, and potentially systematics) was fully determined by the model ID. Following this recipe, we created 100 independent Gaussian training sets and used them to train 100 Gaussian networks. Repeating this procedure with 100 \srolltwo{} training sets, we retrained the set of 100 Gaussian networks to obtain 100 \srolltwo-retrained networks. Using a single test map with input $\tau=0.06$, we find $\sigma(\Delta_{\rm NN})\simeq0.0024$ for Gaussian NN models tested on Gaussian maps, and $\sigma(\Delta_{\rm NN})\simeq0.0034$ for minimally retrained NN models tested on \srolltwo{}. In both cases this represents about $40\%$ of the corresponding value of $\sigma(\tau_{\rm NN})$ reported in Tables~\ref{tab:gaussian_training} and \ref{tab:sroll2_training}, respectively. 

We can reduce the impact of the NN stochasticity by taking, for each test map, the ensemble average of the $\tau$ estimates over the 100 trained NNs. By doing so, for the case with $f_{\rm sky}=0.5$ and input $\tau=0.06$, we find $\sigma(\tau_{\rm NN})\simeq 0.0054$ for Gaussian models applied to Gaussian maps and $\sigma(\tau_{\rm NN})\simeq 0.0083$ for retrained models applied to \srolltwo{} simulations.

We also evaluated the correlation coefficient between the predictions of pairs of models $(j,k)$, tested on the same 10,000 simulations, for both Gaussian and \srolltwo{} training and testing, respectively. In both cases, we find $\rho_{jk}\simeq 0.84$, in agreement with what is expected if Eq.~\eqref{eq:NNerror_model} holds and the models' epistemic errors are uncorrelated, ${\rm Cov}(\Delta^j_{\rm NN},\Delta^k_{\rm NN})=\delta^{\mathcal{K}}_{jk}\sigma^2(\Delta_{\rm NN})$. In the following section we describe how we applied our CNN models to \planck{} maps to infer the value of $\tau$ from data, estimating its uncertainty from simulations and using the ensemble average over 100 trained models to reduce the impact of the NN stochasticity.

\section{Results on \planck{} data}\label{sec:planck}

\begin{figure*}
    \centering
    \includegraphics[width=\linewidth]{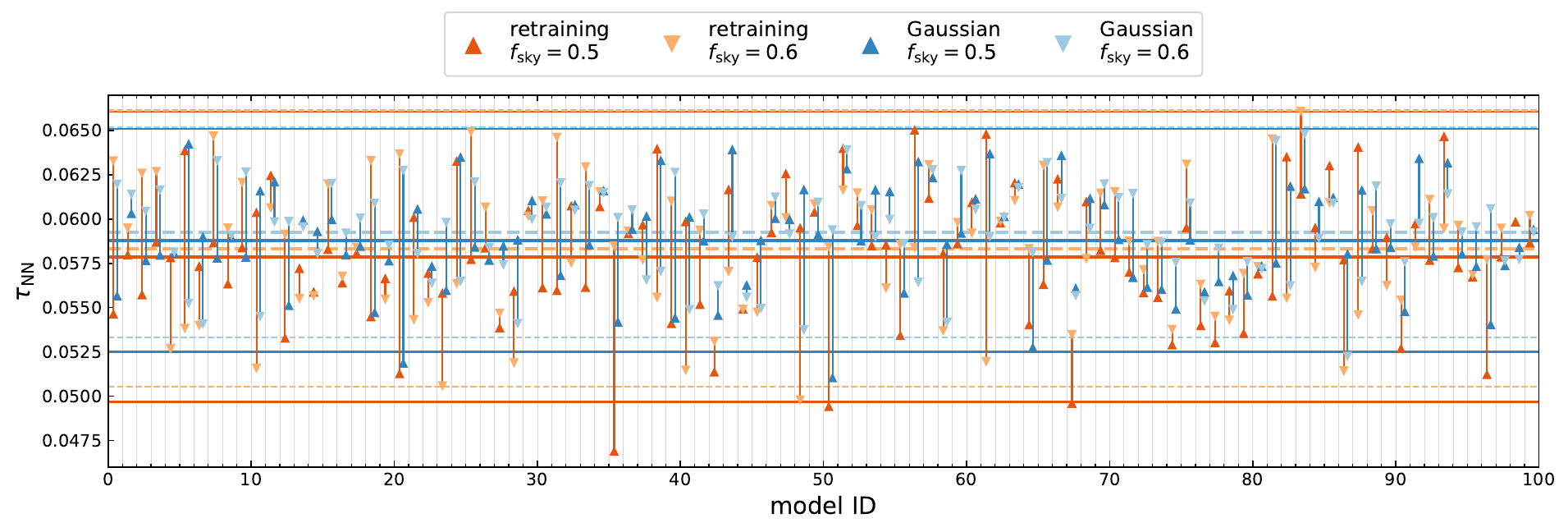}
    \caption{
    NN predictions of $\tau$ from \planck{} 100+143~GHz data, resulting from training 100 equivalent models with different random initial weights and random seeds for training data, considering Gaussian two-channel training (blue tones) versus \srolltwo{} retraining (orange tones), and $f_{\rm sky}=0.5$ (downward triangles) versus $f_{\rm sky}=0.6$ (upward triangles). Colored triangle markers show the best-fit values for the single models and horizontal lines in the corresponding colors show the ensemble average of $\tau_{\rm}$ (middle) $\pm$ the $68\%$ confidence interval (upper and lower lines). \label{fig:planck_nn_variation}}
\end{figure*}

\begin{figure}
    \centering
    \includegraphics[width=\columnwidth]{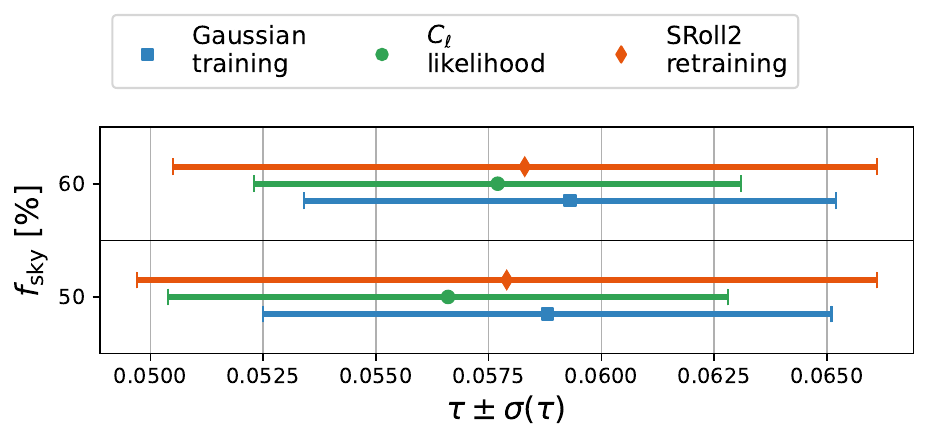}
    \caption{Results on $\tau$ obtained from \planck{} \srolltwo{} data. The values in this plot are shown in Table~\ref{tab:planck_results_masks}.} \label{fig:planck_masks}
\end{figure}

\begin{table*}

\centering
\caption{Results from \planck{} data on two different sky masks, using Gaussian NNs, \srolltwo-retrained NN models, and the empirical $C_\ell$-based likelihood presented in \cite{2020A&A...635A..99P}. The NN results are averaged over 100 models, and $\sigma(\tau_{\rm NN})$ is computed from 10,000 simulations with input $\tau=0.058$.} \label{tab:planck_results_masks}
    \begin{tabular}{ccccccc}
    \hline\hline
     \noalign{\smallskip}
    \multicolumn{7}{c}{Predictions on \planck{} \srolltwo{} data}\\
     \noalign{\smallskip}
    \hline
    \noalign{\smallskip}
     &  \multicolumn{2}{c}{143+100~GHz} & \multicolumn{2}{c}{143+100~GHz}  & \multicolumn{2}{c}{143x100~GHz}\\
     &  \multicolumn{2}{c}{Gaussian training} & \multicolumn{2}{c}{\srolltwo{} retraining} & \multicolumn{2}{c}{$C_\ell$ likelihood} \\
     \noalign{\smallskip}
    \hline
    \noalign{\smallskip}
    $f_{\rm sky}$ & $\tau_{\rm NN}$ & $\sigma(\tau_{\rm NN})$ & $\tau_{\rm NN}$ & $\sigma(\tau_{\rm NN})$ & $\tau$ & $\sigma(\tau)$ \\
    \noalign{\smallskip}
    $50\%$ & 0.0588 & 0.0063 & 0.0579 & 0.0082 & 0.0566 & 0.0062 \\ 
    $60\%$ & 0.0593 & 0.0059 & 0.0583 & 0.0078 & 0.0577 & 0.0054 \\
    \hline\hline
    \end{tabular}
\end{table*}

As shown in Sects.~\ref{ssec:retraining} and \ref{ssec:error_bars}, by retraining on the \srolltwo{} simulations, we are able to obtain a CNN-based model that yields unbiased results on unseen \srolltwo{}  test simulations generated with fixed $\tau\in\{0.05,\, 0.06, \, 0.07\}$. Having thus confirmed the robustness of our method, we moved to real \planck{} data and proceeded to predict $\tau$ from the 100 and 143~GHz \srolltwo{} HFI maps.

We obtained our baseline $\tau$ estimate by taking the average of the inferred values from the 100 minimally retrained NNs applied to \planck{} data for a sky mask with $f_{\rm sky}=0.5$, resulting in a mean estimate of $\tau_{\rm NN}\simeq0.0058$. Figure~\ref{fig:planck_nn_variation} shows the obtained $\tau$ values for each of these NN models. Following the conclusions of the previous sections, since the learned $\sigma_{\rm NN}(\tau)$ is inadequate as an error prediction, we estimated the uncertainty from simulations. In practice, we generated a set of 10,000 \srolltwo{} simulations realized with $\tau=0.058$ and average the $\tau_{\rm NN}$ estimates over 100 networks. Afterwards, we computed the standard deviation over 10,000 simulations. Our final inference on \planck{} maps in this baseline case results in:
\begin{equation}
\tau_{\rm NN} = 0.0579 \pm 0.0082 \quad (\planck{}\, {\rm 100}+{\rm 143\, GHz})\, .
\end{equation}
This value is in very good agreement with the $\tau$ estimates obtained with an empirical likelihood based on cross-QML power spectra, presented in \cite{2020A&A...635A..99P} (hereafter P2020) applied to the same \planck{} maps and constructed from the same \srolltwo{} simulations that we use in this work. In particular, P2020 obtained $\tau=0.0566^{+0.0053}_{-0.0062}$ on the $f_{\rm sky}=0.5$ sky mask. We note that the uncertainty from our NN method is $\sim 30\%$ larger. As previously described, this is due to the fact that our NN estimator does not reach minimum variance and that we relied on the retraining strategy leading to larger errors. However, the fact that we obtain a $\tau$ value in agreement with the literature while using an inherently different inference approach that is, for the first time, fully based on NNs, represents a remarkable result of this work. 

We also applied the Gaussian NN model to \planck{} data, deriving the best-fit parameter value and error bars analogously. We note that, although the Gaussian model leads to results that are mildly biased by up to $\sim 0.5\sigma$ when applied to \srolltwo{} maps with low CMB input signal ($\tau=0.05$), the bias is below $0.15\sigma$ when $\tau=0.06$, as displayed in the fifth column of Table~\ref{tab:gauss_training_sr_test}. In this case, using the same $f_{\rm sky}=0.5$ mask, we obtained $\tau_{\rm NN}=0.0588\pm 0.0063$. The statistical uncertainty is lower for this second method, as we omitted retraining on systematics, and similar to the one obtained from the empirical likelihood presented in P2020. 

Lastly, as a robustness test, we applied these same methods to a second sky mask, with a larger sky coverage of $f_{\rm sky}=0.6$. The parameter estimated remained stable for both the retrained and the Gaussian model, with lower uncertainties. The NN predictions of the single models on $f_{\rm sky}=0.6$ are displayed in Fig.~\ref{fig:planck_nn_variation}. A summary of our results on \planck{} maps is shown in Fig.~\ref{fig:planck_masks} and Table~\ref{tab:planck_results_masks}.

\section{Conclusions}\label{sec:conclusion}

In this paper, we present the first cosmological parameter inference on \planck's CMB polarization maps that is performed entirely by neural networks. We estimated the optical depth to reionization, $\tau$, from the \srolltwo{} low resolution polarization maps of \planck-HFI  at 100 and 143~GHz. These maps are known to contain a significant level of residual systematic effects at large angular scales that, if ignored, would bias cosmological results. These spurious signals are non-Gaussian and hard to model in an analytical way. For this reason, in the literature \citep[P2020]{2020A&A...635A..99P}, the estimation of $\tau$ from these maps is obtained by sampling an empirical $EE$ cross-spectrum likelihood \citep{Planck:2019nip,2020FrP.....8...15G}, built from a set of  realistic \srolltwo{} simulations \citep{Delouis:2019bub}. 

In this work, we approached this problem through NN-based inference applied directly on the map domain. One of the benefits of this method is that it does not require an analytical model of the data but, instead, relies solely on simulations to train a regression model. In particular, we used the \nnhealpix{} algorithm to build our NN models, allowing the application of convolutional layers on the sphere. We considered several setups to train and validate CNNs on multiple sets of simulations, before applying them to \planck{} data. We adopted the moments loss function \citep{Jeffrey:2020itg} to learn the mean and standard deviation of the marginal posterior on $\tau$ inferred from Stokes $Q$ and $U$ maps pixelized on a grid at $\nside=16$ ($\sim 4^\circ$). To find the best training method, we started from simulations of a single frequency channel of CMB with coadded Gaussian correlated noise and, step by step, moved to more complex setups that involved two frequency channels containing CMB, noise, and systematic effects. We compared the results obtained with NNs with the ones from a standard Bayesian method that applies the HL likelihood to $EE$ cross-spectra. Our main results and conclusions from the analysis applied to simulations are the following:

\begin{enumerate}
\item When trained and applied to Gaussian simulations, the NN models are able to retrieve unbiased values of $\tau$ directly from maps. Additionally, by using the moments loss function reported in Eq.~\eqref{eq:loss_function}, the models can also learn and return an error estimate that is consistent with the spread of the best-fit values on the test set. 
\item When trained using maps from two frequency channels that share the same cosmological signal, the NNs are able to effectively combine the information from both maps. This leads to improved accuracy in the $\tau$ estimates and smaller uncertainties. This ability to combine information from different channels is a key advantage of the NN approach as, in the future, it would allow for a straightforward combination of all available data sets without the need for a joint model, thus reducing the impact of noise and systematics.
\item A comparison of the NN estimates with the ones obtained from the HL cross-spectrum method applied to Gaussian simulations shows that the NN approach leads to higher uncertainties by about 20\%. This implies that the NN estimator, although unbiased, does not reach the minimum variance. In order to further improve the performance of the estimator, future work should focus on optimizing the spherical convolution algorithm, the model architecture, and the training procedure. This will help to minimize the uncertainties and reach the best possible performance.
\item The application of the Gaussian two-channel model to the \srolltwo{} simulations, which include systematic effects, leads to inaccurate estimates on $\tau$, as does the use of the HL likelihood. Although expected, this observed bias is much smaller (nearly unbiased for $\tau\sim0.06$) than that seen for the single-channel model, demonstrating that the neural network is able to identify common features in the maps, efficiently ignoring the uncorrelated signal between different channels. 
\item To recover fully unbiased results on \srolltwo{} maps, as a prerequisite to apply our NN model to \planck{} data, we needed to train NNs on maps that incorporate instrumental systematic effects. Due to the limited number of available \srolltwo{} simulations, we adopted a minimal retraining approach, building on the good results already obtained with the Gaussian models. This approach helps to minimize the issue of overfitting, but it also leads to slightly larger errors in the recovered $\tau$ values. 
\item In more complex scenarios, when we applied the NN models to the \srolltwo{} maps, we found that the error estimate learned by the NN, $\sigma_{\rm NN}(\tau)$, underestimated the spread evaluated on the empirical distribution of the test maps, $\sigma(\tau_{\rm NN})$. This suggests that the NN model did not capture the full range of uncertainty in the data. To overcome this issue, we proceeded by evaluating the final error on $\tau$ through simulations, by taking the ensemble average of 100 NN models. This helped to reduce the impact of the epistemic uncertainty caused by the intrinsic stochasticity of the NN estimator. 
\end{enumerate}

After evaluating and validating the performance of the NNs on simulations, we applied our trained models to \planck{} \srolltwo{} maps at 100 and 143~GHz. For the minimally retrained model, which is the one that leads to fully unbiased results on the \srolltwo{} simulations, we obtain $\tau_{\rm NN}=0.0579\pm0.0082$ on our fiducial $f_{\rm sky}=0.5$ mask. This value is in very good agreement with the one obtained from the empirical likelihood based on cross power spectra reported in P2020, which relies on the same set of simulations. We consider this a remarkable result of our work, given the fact that the two estimators are intrinsically different. However, we note that our final uncertainty on the $\tau_{\rm NN}$ estimate, which we evaluated from simulations and the ensemble average of 100 NN models, is about $30\%$ larger than the one obtained in P2020. This is because our NN estimator does not reach minimum variance and, moreover, we could rely only on a limited number of \srolltwo{} simulations to inform the NN about systematic effects. The minimal retraining approach allows us to achieve unbiased results, but at the cost of an increased variance.

An effective robustness test against systematics-induced ``unknown unknowns'' is to predict $\tau$ from \srolltwo{} simulations using the Gaussian network (as described in Sect. \ref{sec:gaussian_training}), which can be considered agnostic to the strong non-Gaussian map features characteristic for \srolltwo{} maps. Given its good performance on \srolltwo{} simulations for $\tau\sim0.06$, we applied the Gaussian model to the \planck{} data. In this second case we obtain $\tau_{\rm NN}=0.0588\pm0.0063$, in agreement with the estimate reported in the literature, and with a similar level of uncertainty. 

As an additional robustness test of the NN approach, we considered a second mask that retains a larger sky fraction of $f_{\rm sky}=0.6$ and find consistent results. The summary of our results is reported in Table~\ref{tab:planck_results_masks} and Fig.~\ref{fig:planck_masks}, showing full stability of the retrieved $\tau_{\rm NN}$ estimations.

Concluding, in this work we present a first thorough application of NN-based inference to real CMB maps. It is important to stress that obtaining reliable results on real data required a significant effort to validate and test our models on different setups and to develop training strategies that can effectively cope with systematic effects. This highlights the fact that NN models developed to perform well on simplified simulations cannot always be straightforwardly applied to real data and need careful consideration of the training and validation procedures.
Nonetheless, the consistent and robust results we obtain demonstrate that NNs represent a promising tool that could complement standard statistical data analysis techniques for CMB observations, especially in cases where the Gaussian CMB signal is contaminated by spurious effects that cannot be analytically described in a likelihood model. This is particularly relevant for the ongoing search for primordial gravitational waves, constrained by large-scale $B$-modes which are targeted by a number of near-future experiments such as the \textit{Simons Observatory} \citep{SimonsObservatory:2018koc}, \textit{LiteBIRD} \citep{LiteBIRD:2022cnt} and \textit{CMB-S4} \citep{Abazajian:2019eic}. However, additional optimization and validation of this approach must be developed before tackling this challenge.

\begin{acknowledgements}
The authors acknowledge financial support from the COSMOS network (www.cosmosnet.it) through the ASI (Italian Space Agency) Grants 2016-24-H.0 and 2016-24-H.1-2018, as well as 2020-9-HH.0 (participation in LiteBIRD phase A). LP acknowledges financial support and computing resources at CINECA provided by the INFN InDark initiative. We acknowledge the use of  \texttt{CAMB} \citep{Lewis:1999bs}, \texttt{healpy} \citep{Zonca2019}, \nnhealpix{} \citep{2019A&A...628A.129K}, \texttt{numpy} \citep{Harris:2020xlr}, \texttt{matplotlib} \citep{Hunter:2007ouj}, and \texttt{keras} \citep{chollet2015keras} software packages. This research used resources of the National Energy Research Scientific Computing Center (NERSC), a U.S. Department of Energy Office of Science User Facility located at Lawrence Berkeley National Laboratory, operated under Contract No. DE-AC02-05CH11231.
\end{acknowledgements}

\bibliographystyle{aa}
\bibliography{bibliography}

\end{document}